\documentstyle[12pt]{article}

\newcommand{\be}{\begin{equation}}
\newcommand{\ee}{\end{equation}}

\newcommand{\bea}{\begin{eqnarray}}
\newcommand{\eea}{\end{eqnarray}} 
\newcommand{\lp}{\left(}
\newcommand{\rp}{\right)}

\begin{document}
\title{Dynamical Manifestation of the Goldstone Phenomenon at 1-loop} 

\author{{Sz. Bors\'anyi$^{1}$, A. Patk\'os$^{2}$ and Zs. Sz\'ep$^{3}$}\\
{Department of Atomic Physics}\\
{E\"otv\"os University, Budapest, Hungary}\\
}
\vfill
\footnotetext[1]{mazsx@cleopatra.elte.hu}
\footnotetext[2]{patkos@ludens.elte.hu}
\footnotetext[3]{szepzs@hercules.elte.hu}
\maketitle
\begin{abstract}
We have calculated the damping rate $\Gamma (|{\bf k}|)$ for classical
on-shell Goldstone modes of the $O(2)$ symmetric scalar fields propagating 
in a thermal medium of the broken symmetry phase taking into account
the effect of the explicit symmetry breaking. The result of the one-loop
analysis can be expanded around $\Gamma (0)$, which depends non-analytically
on the parameter of the explicit symmetry breaking, $h$. $\Gamma (0)$
 vanishes when
$h\rightarrow 0$, demonstrating in this way the absence of the restoring 
force, when the equilibrium direction of the symmetry breaking is modulated
homogeneously.
\end{abstract}

The equivalence of the vacuum states in a system with spontaneous symmetry
breaking should manifest itself also in the dynamical evolution of 
fluctuations and external signals: when an external signal superimposed on
the actual vacuum realizes the transition over to an equivalent vacuum state
no relaxation to the initial vacuum should occur. In this note we present
an explicit analysis of the near equilibrium, linear relaxation 
phenomenon in an $O(2)$-symmetric
scalar field theory from this point of view. It is convenient (and also
more realistic) to study the problem in a system with additional explicit 
breaking of the symmetry, and follow what happens when its strength goes to 
zero. We are going to derive an explicit formula for the decay rate of a 
homogeneous Goldstone modulation of the vacuum displaying the expected 
behaviour in a rather non-trivial way.

Our investigation is an extension of our recent detailed analysis of the
one-loop finite temperature dynamics of the $O(N)$-symmetric scalar
field theories \cite{jako99a}. (See also \cite{tytgat96,boyan99}.) 
The detailed formalism presented there
permits us to restrict the presentation to the most important relations,
since the formal details are very similar to the case with zero explicit
symmetry breaking. We have specialized the discussion of the 
{\it linear response function} to the $O(2)$ case
just to simplify some formulae. The conclusions appear to be of 
intrinsic validity for the dynamical effect of the Goldstone phenomenon.

The study of the Goldstone-propagation through thermal medium is of
particular interest for the interpretation of the pion-sigma dynamics in
heavy ion collisions. Several recent field theoretical studies have dealt
with this problem \cite{boyan95,rischke98,csernai99} mainly concentrating
on the evolution of the heavy (sigma or Higgs) component. Though it has 
been observed in \cite{rischke98} that homogeneous pion condensates do not
decay, the origins of this statement have not been fully explored. The
effect of explicit symmetry breaking was taken into account by some authors
by treating the pions as effectively massive degrees of freedom, without
exploiting the consequences of the approximate symmetry 
\cite{boyan96,mrow97}.

The Lagrangian of the $O(2)$ symmetric scalar field theory with explicit 
symmetry breaking is the following:
\be
L={1\over 2}\lp\partial_\mu\phi_1\rp^2+{1\over 2}\lp\partial_\mu\phi_2\rp^2
-{1\over 2}\mu^2\lp\phi_1^2+\phi_2^2\rp-{\lambda\over
24}\lp\phi_1^2+\phi_2^2\rp^2
+h\phi_1.
\ee
To account for the symmetry breaking one shifts the $\Phi_1$-field:
\be
\phi_1\rightarrow\bar\Phi +\phi_1,
\ee
and we assume that $\bar\Phi$ is given by its equilibrium value.
Then we are led to the following shifted Lagrangian:
\bea
L&=&{1\over 2}\lp\partial_\mu\phi_1\rp^2
+{1\over 2}(\partial_\mu\phi_2)^2
-{1\over 2}m_1^2\phi_1^2
-{1\over 2}m_2^2\phi_2^2
+\phi_1\lp h-\mu^2\bar\Phi-{\lambda\over 6}\bar\Phi^3\rp\nonumber\\
&-&{\lambda\over 24}\lp 4\bar\Phi\phi_1^3+\phi_1^4+4\bar\Phi\phi_1\phi_2^2+
2\phi_1^2\phi_2^2+\phi_2^4\rp
\eea
with the notations
\be
m_1^2=\mu^2+{\lambda\over 2}\bar\Phi^2,\qquad m_2^2=\mu^2+{\lambda\over 6}
\bar\Phi^2.
\ee
Next we sketch the steps followed in the derivation of the effective equations 
for the propagation of a non-thermal signal on a thermalised background:

1. Decomposition of the fields $\phi_i$ into the sum of high-frequency, 
thermalised ($\varphi_i$) and low-frequency, non-thermal ($\Phi_i$) components;

2. Derivation of the equations of motion for $\Phi_i$ in the background
of the thermal components;

3. Averaging the equations over the thermal background, retaining only 
contributions from the two-point functions of the thermalised fields;

4. Approximating the two-point functions resulting from step 3 by 
expressions with at most linear functional dependence on the non-thermal 
fields, what is sufficient for the calculation of the linear response 
function of the theory:

\bea
&
\langle\varphi_i(x)\varphi_j(y)\rangle\approx
\left\langle\varphi_i(x)\varphi_j(y)\rangle \right|_{\Phi=0}+
\left. \displaystyle\int dz\ 
\frac{\delta \langle\varphi_i(x)\varphi_j(y)\rangle}{\delta\Phi_l(z)}
\right|_{\Phi=0}\cdot\Phi_l(z)\equiv\nonumber\\
&
\langle\varphi_i(x)\varphi_j(y)\rangle^{(0)}
+\langle\varphi_i(x)\varphi_j(y)\rangle^{(1)}.
\eea

The first ($\Phi_i$-independent) term is different from zero only for
$i=j$ and it is given by its equilibrium expression,
with thermal population restricted to the $p_0>\Lambda$ range:
\bea
&
\langle\varphi_i\varphi_j\rangle^{(0)}=\delta_{ij}
\displaystyle\int\frac{d^4p}{(2\pi )^4}
2\pi\delta (p^2-M_i^2)(\Theta (p_0)+\tilde n(|p_0|)),\nonumber\\
&
\tilde n(x)=\Theta (x-\Lambda ){1\over e^{\beta x}-1}.
\eea
In the equilibrium distribution one makes use of the physical masses of the 
fields, to be deduced below (see Eq.(\ref{masses})).

The linearised equations of motion resulting from the above described 
procedure are the following:
\bea
\lp\partial^2+m_1^2+{\lambda\over 2}\langle\varphi_1^2\rangle^{(0)}
+{\lambda\over 6}\langle\varphi_2^2\rangle^{(0)}\rp\Phi_1&=&
-{\lambda\over 2}\bar\Phi\langle\varphi_1^2\rangle^{(1)}-{\lambda\over 6}
\bar\Phi\langle\varphi_2^2\rangle^{(1)},\nonumber\\
\lp\partial^2+m_2^2+{\lambda\over 2}\langle\varphi_2^2\rangle^{(0)}
+{\lambda\over 6}\langle\varphi_1^2\rangle^{(0)}\rp\Phi_2&=&
-{\lambda\over 3}\bar\Phi\langle\varphi_1\varphi_2\rangle^{(1)}.
\label{eqsmot}
\eea
In addition the equation for the equilibrium (static)
value of the vacuum expectation, $\bar\Phi$ is also found:
\be
m_2^2\bar\Phi -h+{\lambda\over 2}\bar\Phi\langle\varphi_1^2\rangle^{(0)}+
{\lambda\over 6}\bar\Phi\langle\varphi_2^2\rangle^{(0)}=0.
\ee
Below for the sake of simpler notation, we shall use the following
conventional abbreviation:
\be
\langle\varphi_i(x)\varphi_j(y)\rangle \equiv\Delta_{ij}(x,y).
\ee

In order to proceed with the discussion of the decay rate of the Goldstone
 condensate we have to derive an equation for 
 $\langle\varphi_1\varphi_2\rangle^{(1)}$, which determines the
propagation of $\Phi_2$:
\be
\lp\partial_x^2-\partial_y^2+{\lambda\over 3}\bar\Phi^2\rp
\Delta_{12}^{(1)} (x,y)=
-{\lambda\over 3}\bar\Phi \left(\Phi_2(x)
\Delta_{22}^{(0)}(x-y)-\Phi_2(y)\Delta_{11}^{(0)}(x-y)\right).
\ee

After performing the combined Wigner-Fourier transformation 
\cite{heinz86,mrow90,jako99a} of this
equation one has
\be
\lp 2p\cdot k +{\lambda\over 3}\bar\Phi^2\rp\Delta_{12}^{(1)}(k,p)=
-{\lambda\over 3}\bar\Phi\Phi_2(k)\lp\Delta_{22}^{(0)}\lp p+k/2\rp-
\Delta_{11}^{(0)}\lp p-k/2\rp\rp.
\ee
Finally, the effective linear wave equation for the field $\Phi_2$ looks like:
\bea
&
\lp-k^2+{h\over\bar\Phi}\rp\Phi_2(k)=
{\lambda\over 3}\displaystyle\int \frac{d^4p}{(2\pi )^4}
\lp\Delta_{11}^{(0)}(p)-
\Delta_{22}^{(0)}(p)\rp\Phi_2(k)\nonumber\\
&
+\lp\frac{\lambda}{3}\bar\Phi \rp^2\Phi_2(k)
\displaystyle\int\frac{d^4p}{(2\pi )^4}
{1\over 2p\cdot k+{\lambda\over 3}\bar\Phi^2}\lp\Delta_{22}^{(0)}(p+k/2)-
\Delta_{11}^{(0)}(p-k/2)\rp.
\label{eff_eq}
\eea

For $k=0$ the right hand side of this equation vanishes, what reflects
the effect of the Goldstone-theorem on the self-energy function of
$\Phi_2$. A similar analysis of the first equation of motion in 
Eq.(\ref{eqsmot}) shows that the diagonal thermal
two-point functions appearing on the right hand side do not
contribute any local term, which could be interpreted as the correction of 
the mass-square appearing on the left hand side. Therefore
\be
M_H^2={\lambda\over 3}\bar\Phi^2+{h\over\bar\Phi},\qquad
M_G^2={h\over \bar\Phi}.
\label{masses}
\ee

The decay rate of the $\Phi_2$-field is determined by the imaginary part of 
the self-energy function appearing on the right hand side of (\ref{eff_eq}):
\be
({\lambda\over 3}\bar\Phi)^2{\rm Im}\left[ \int \frac{d^4p}{(2\pi )^4}
\frac{1}{2p\cdot k+M_H^2-M_G^2}\lp\Delta_{22}^{(0)}(p+k/2)-
\Delta_{11}^{(0)}(p-k/2)\rp\right].
\ee
The kinematical analysis of the range of variables contributing to 
this integral is much more involved than in the case when the symmetry
breaking happens exclusively spontaneously (see \cite{boyan96} for
a detailed discussion). After the straightforward but 
tedious procedure we find for the imaginary part of the self-energy function:
\bea
&&{\rm Im\ }\Pi_2(k_0,|{\bf k}|)=\lp{\lambda\over 3}\bar\Phi
\rp^2\frac{1}{16\pi |{\bf k}|}\times\nonumber\\
&&\left[\Theta (-k^2)
\int\limits_{\alpha_{-}}^\infty ds\ \left(n(s)-n(s-k_0)\right)
+\Theta (-k^2)
\int\limits_{\alpha_{+}}^\infty ds\ \left(n(s+k_0)-n(s)\right)\right.
\nonumber\\
&+&\Theta (k^2)\Theta (M_G^2-M_H^2)\Theta \lp(M_H-M_G)^2-k^2\rp
\int\limits_{\alpha_{-}}^{\alpha_{+}}ds~(n(s)-n(s-k_0))\nonumber\\
&+&\Theta (k^2)\Theta (M_H^2-M_G^2)\Theta \lp(M_G-M_H)^2-k^2\rp
\int\limits_{\alpha_{+}}^{\alpha_{-}}ds\ \lp n(s+k_0)-n(s)\rp\nonumber\\
&-&\left.\Theta (k^2)\Theta \lp k^2-(M_H+M_G)^2\rp
\int\limits_{\alpha_{-}}^{\alpha_{+}}ds\ \lp 1+n(s)+n(k_0-s)\rp\right],
\eea
with
\be
\alpha_{\pm}=\left|
\frac{1}{2k^2}\left(k_0\left(k^2-M_H^2+M_G^2\right)\pm
|{\bf k}|\sqrt{(k^2-M_H^2+M_G^2)^2-4k^2M_G^2} \right)\right|.
\ee

In case of on-shell propagation ($k_0^2-|{\bf k}|^2=M_G^2$) and
$M_G<M_H$, only the fourth
term contributes. With its help for the decay rate one finds
\be
\Gamma ={(\lambda\bar\Phi /3)^2\Theta(M_H-2M_G)\over
32\pi|{\bf k}|\sqrt{|{\bf k}|^2+M_G^2}}
\left[ \int\limits_{\alpha_{+}}^{\alpha_{+}+k_0}ds\ n(s)
-\int\limits_{\alpha_{-}}^{\alpha_{-}+k_0}ds\ n(s) \right].
\label{rate}
\ee

For $M_G=0$ and $k_0\neq 0$ it reproduces our recent result for the
damping rate of a Goldstone wave without explicit symmetry breaking 
\cite{jako99a}:
\be
\left.\Gamma ({\bf k})\right|_{M_G=0}=\left(\frac{\lambda}{3}\bar\Phi \right)^2
{1\over 32\pi |{\bf k}|}~n\lp{M_H^2\over 4|{\bf k}|})\rp.
\ee
For $M_G\neq 0$, the damping rate is a continuous function of $|{\bf k}|^2$
and can be expanded around $|{\bf k|}=0$:
\bea
&\Gamma (|{\bf k}|=0)=&{\lambda^2\bar\Phi^2\over 288\pi }
\Theta \lp M_H-2M_G\rp
\frac{M_H}{M_G^3}\sqrt{M_H^2-4M_G^2}\times\nonumber\\
&&
\lp\exp \lp-\frac{\beta M_H^2}{2M_G}\rp 
-\exp \lp -\frac{\beta (M_H^2+M_G^2)}{2M_G}\rp
\rp\times\nonumber\\
&&
\lp\exp \lp-\frac{\beta (M_G^2+M_H^2)}{2M_G}\rp-1\rp^{-1}
\lp\exp \lp-\frac{\beta M_H^2}{2M_G}\rp-1\rp^{-1}.\nonumber\\
&
\label{ratexp}
\eea
In particular, it can be evaluated for $T \gg M_G,M_H$, with a result which
exactly coincides with the classical expression for the decay rate. 
This latter arises from Eq.(\ref{rate}) with the replacement $n(s)\rightarrow
n_{cl}(s)=T/s$. It can be derived from the classical evolution equations of 
the $O(2)$ symmetric scalar model as explained in \cite{buch97}.

However, for $M_G\rightarrow 0$ the expression (\ref{ratexp}) does not follow 
the classical theory, but
vanishes non-analytically as $\exp (-\beta M_H^2/2M_G)$. This result 
demonstrates that not only the self-energy function, but {\it also the decay 
rate vanishes for $|{\bf k}|=0$, when the strength of the explicit symmetry 
breaking  goes to zero}.

The quantity $\Gamma (h, |{\bf k}|)$ arises at this order of the perturbation
theory as the difference of the rates of two processes: the transformation of 
the Goldstone-wave into a Higgs-wave with the absorption of a 
(high-frequency) thermal Goldstone-particle and its inverse.
From the macroscopic point of view our result presents the rate of
transformation of a Goldstone-signal into Higgs-signal when propagating 
through the thermalised medium. In the approximation when one assumes
that a single act of transformation is not followed by any further
interaction with the thermal bath, it gives also the inverse life-time of the
Goldstone-wave. However, if one anticipates further multiple transformations
between the Goldstone and the Higgs forms of propagation, 
then the damping rate of 
the original Goldstone signal will considerably increase for small, but finite
$h$ due to the more efficient damping of the $k_0\approx T$ Higgs-waves.
However, for $h=0$ the full rate still vanishes, since the 
Goldstone-to-Higgs 
transformation rate itself is zero. A recent analysis of A. Jakov\'ac
\cite{jako99b} presents a systematic way for taking into account the effect
of the larger Higgs-width in the Goldstone-propagator.

The authors acknowledge valuable discussions with A. Jakov\'ac, 
S. Mr\'ow\-czy\'n\-sky and P. Petreczky. This work has been supported by the 
research contract OTKA-T22929.


\begin{thebibliography}{9}
\bibitem{jako99a} A. Jakov\'ac, A. Patk\'os, P. Petreczky and Zs. Sz\'ep, 
{\it Effective theory for soft fluctuation modes in the spontaneously 
broken phase of the N-component scalar field theory} hep-ph/9905439
(to appear in the November 15 issue of Phys. Rev. {\bf D})

\bibitem{tytgat96} R. Pisarski and M. Tytgat, Phys. Rev. {\bf D54}
(1996) 2989

\bibitem{boyan99} D. Boyanovsky, H.J. de Vega and S.-Y. Wang,
{\it Dynamical Renormalisation Group Approach to Quantum Kinetics in Scalar
and Gauge Theories}, hep-ph/9909369

\bibitem{boyan95} D. Boyanovsky, H.J. de Vega, R. Holman, D.S. Lee and 
A. Singh, Phys. Rev. {\bf D51} (1995) 4419

\bibitem{rischke98} D. Rischke, Phys. Rev. {\bf C58} (1998) 2331

\bibitem{csernai99} L.P. Csernai, P.J. Ellis, S. Jeon and J.I. Kapusta,
{\it Dynamical evolution of the scalar condensates in heavy ion collisions},
CERN-TH/99-193, NUC-MINN-99/9-T, nucl-th/9908020

\bibitem{boyan96} D. Boyanovsky, M. D'Attanasio, H.J. de Vega and R. Holman,
Phys. Rev. {\bf D54} (1996) 1748

\bibitem{mrow97} S. Mr\'owczy\'nski, Phys. Rev. {\bf D56} (1997) 2265

\bibitem{heinz86} U. Heinz, Ann. Phys. (N.Y.) {\bf 168} (1986) 148

\bibitem{mrow90} P. Danielewicz and S. Mr\'owczy\'nski, Nucl. Phys. {\bf B342} 
(1990) 345
\bibitem{buch97} W. Buchm\"uller and A. Jakov\'ac, Phys. Lett. {\bf B407} 
(1997) 39
\bibitem{jako99b} A. Jakov\'ac, (private communication)
\end{thebibliography}
\end{document}